\documentclass [11pt,a4paper]{article}
\usepackage{jheppub}

\usepackage{hyperref}%[breaklinks]
\usepackage[latin1,utf8]{inputenc}
\usepackage{graphicx}
\usepackage{color}
\usepackage{amsfonts,amsthm,amssymb}
\usepackage{bm,bbm}
\usepackage[OT2,OT1]{fontenc}
\usepackage{tikz}

\newcommand\cyr{%
\renewcommand\rmdefault{wncyr}%
\renewcommand\sfdefault{wncyss}%
\renewcommand\encodingdefault{OT2}%
\normalfont
\selectfont}
\DeclareTextFontCommand{\textcyr}{\cyr}

\newcommand{\be}{\begin{equation}}
\newcommand{\ee}{\end{equation}}
\newcommand{\ba}{\begin{eqnarray}}
\newcommand{\ea}{\end{eqnarray}}

\renewcommand{\texttt}{{}}

\def\bs{\begin{subequations}}
\def\es{\end{subequations}}
\def\a{\alpha}

\def\de{\delta}

\def\ve{\varepsilon}

\def\om{\omega}

\def\s{\sigma}

\def\cK{\mathcal{K}}

\def\cO{\mathcal{O}}

\def\ds{d_{\rm S}}
\def\dh{d_{\rm H}}

\def\p{\partial}

\newcommand{\Eq}[1]{(\ref{#1})}
\def\com{\color{magenta}}
\def\cob{\color{blue}}

\newcommand{\oarX}[1]{\href{http://arxiv.org/abs/#1}{{\ttfamily\com arXiv:#1}}}
\newcommand{\arX}[1]{\href{http://arxiv.org/abs/#1}{{\ttfamily\com arXiv:#1}}}
\newcommand{\doin}[6]{\href{http://dx.doi.org/#1}{{\cob {\it #2} {\bf #3 #4} (#6) #5}}}
\newcommand{\doinn}[5]{\href{http://dx.doi.org/#1}{{\cob {\it #2} {\bf #3} (#5) #4}}}
\newcommand{\doij}[5]{\href{http://dx.doi.org/#1}{{\cob {\it #2} {\bf #3} (#5) #4}}}

\newcommand{\procsinm}[5]{in \emph{#1}, #2 eds., #3, #4 (#5)}

\newcommand{\tia}[1]{\textit{#1},}

\def\lp{\ell_{\rm Pl}}
\def\tp{t_{\rm Pl}}
\def\ep{E_{\rm Pl}}
\def\rme{e}
\def\rmd{d}
\def\rmi{i}

\begin{document}

\title{New Standard Model constraints on the scales and dimension of spacetime}

\author[a]{Andrea Addazi,}
\emailAdd{andrea.addazi@qq.com}
\affiliation[a]{Center for Field Theory and Particle Physics \& Department of Physics, Fudan University, 200433 Shanghai, China}

\author[b]{Gianluca Calcagni,}
\emailAdd{g.calcagni@csic.es}
\affiliation[b]{Instituto de Estructura de la Materia, CSIC, Serrano 121, 28006 Madrid, Spain}

\author[a]{Antonino Marcian\`o}
\emailAdd{marciano@fudan.edu.cn}

\abstract{Using known estimates for the kaon--antikaon transitions, the mean lifetime of the muon and the mean lifetime of the tau, we place new and stronger constraints on the scales of the multi-fractional theories with weighted and $q$-derivatives. These scenarios reproduce a quantum-gravity regime where fields live on a continuous spacetime with a scale-dependent Hausdorff dimension. In the case with weighted derivatives, constraints from the muon lifetime are various orders of magnitude stronger than those from the tau lifetime and the kaon--antikaon transitions. The characteristic energy scale of the theory cannot be greater than $E_*>3\times 10^2\,{\rm TeV}$, and is tightened to $E_*>9\times 10^{8}\,{\rm TeV}$ for the typical value $\a=1/2$ of the fractional exponents in the spacetime measure. We also find an upper bound $\dh<2.9$ on the spacetime Hausdorff dimension in the ultraviolet. In the case with $q$-derivatives, the strongest bound comes from the tau lifetime, but it is about 10 orders of magnitude weaker than for the theory with weighted derivatives.}

\date{October 17, 2018}


\preprint{\doij{10.1007/JHEP12(2018)130}{JHEP}{12}{130}{2018} [\arX{1810.08141}]}

\maketitle

%%%%%%%%%%%%%%%%%%%%%%%%%%%%%%%%%%%%%%%%%%%%%%%%%%%%%%%%%%%%%%%%%%%%%%%%%%%%%%%%%%%%%%%%%
%%%%%%%%%%%%%%%%%%%%%%%%%%%%%%%%%%%%%%%%%%%%%%%%%%%%%%%%%%%%%%%%%%%%%%%%%%%%%%%%%%%%%%%%%

\section{Introduction}
\noindent
The search for signatures of a quantum theory of gravity has been gradually increasing in the last few years, thanks to the theoretical advances in the field and to the new generations of experiments in particle physics, astrophysics and cosmology. Aside from checking specific proposals, there is interest in probing dimensional flow, a feature common to all quantum gravities. This is the change of spacetime dimensionality with the observation scale \cite{tH93,Car09,fra1,revmu,Car17}. The idea that we can observe deviations from the topological dimension $D=4$ dates back to early attempts to constrain toy models in $D=4-\ve$ dimensional regularization from quantum electrodynamics (QED) and cosmology \cite{ScM,ZS,MuS,CO}. It was then revived in more realistic quantum-gravity-related scenarios, multi-scale spacetimes, where geometry is endowed with intrinsic scales and spacetime dimensionality (defined differently from the na\"ive topological dimension $D$) varies continuously from small to large scales; see \cite{revmu} for a review. The presence of just one fundamental length scale $\ell_*$ (inversely proportional to a fundamental energy scale $E_*$) is sufficient to have dimensional flow.

While multi-scale spacetimes can describe generic backgrounds realized by any quantum gravity admitting a continuum limit and realizing dimensional flow as an emergent property, they also host stand-alone theories where dimensional flow is explicit. The phenomenology of these \emph{multi-fractional theories} has been intensely scrutinized. In particular, the Standard Model of electroweak and strong interactions has been constructed both in the so-called theory with weighted derivatives \cite{frc8,frc13} and in the one with $q$-derivatives \cite{frc13,frc12}. In the case with weighted derivatives, constraints on $\ell_*$ and $E_*$ were found from the electron $g-2$ factor, the fine-structure constant and the Lamb shift, while in the case with $q$-derivatives there are no constraints from the fine-structure constant and a rather weak bound from the muon lifetime was considered.

In this paper, we find new Standard Model constraints of these two theories. For both, we will consider kaon-antikaon transitions and the tau lifetime, while for the theory with weighted derivatives we will also get information from the muon lifetime. The latter is the strongest bound to date on the scales of this theory, while in the case with $q$-derivatives astrophysical constraints are more compelling \cite{revmu}. Independently of the value of these bounds for these specific theories, the main point is that particle-physics experiments \emph{can} unravel non-perturbative signatures of quantum gravity, or at least constrain the geometry of spacetime efficiently, contrary to what one would be led to believe by oft-quoted perturbative arguments.

This paper is not the first one to consider $K^0-\bar{K}^0$ processes to test quantum gravity. In \cite{EHNS}, the Hamiltonian contribution $H-H^\dagger\neq 0$ of generic effects violating standard quantum mechanics in the Standard Model were constrained very tightly. The main difference between the seminal analysis of Ref.~\cite{EHNS} and our scenario is in the Hamiltonian mass-matrix sector of the kaon--antikaon system. The authors of \cite{EHNS} considered, in a model-independent way, the case of CPT violating phases, leading to a loss of hermiticity of the mass matrix. This is thought to be induced by quantum-gravity decoherence effects ---for example, from virtual black-hole pairs. On the other hand, in our case the mass matrix remains hermitian, i.e., no CPT violation or quantum decoherence arise from a multi-scale spacetime; see section \ref{sec3}. Thus, $H-H^\dagger=0$ and we evade the strong bound of \cite{EHNS}. In our scenarios we get different observables in the kaon--antikaon mass matrix. For the theory with weighted derivatives, the main effect is encoded into a spacetime-dependence of the Long-time kaon and Short-time antikaon mass difference, while having no CPT violating phases. In the theory with $q$-derivatives, what changes is the relation between the decay rate and the physical lifetime.

Such a crucial difference between the case of \cite{EHNS} and ours is related to the fact that, even if the Lorentz symmetry is broken in multi-scale backgrounds, the CPT symmetry is not violated necessarily. This situation loosely reminds us of other cases, in which CPT invariance is subtly compatible with the simultaneous violation of two of its constituent discrete symmetries.

In section \ref{sec2}, we review multi-fractional theories in a self-contained way and clarify how observables are computed; the only detail of importance we omit is the exact form of the full $SU(3)\times SU(2)\times U(1)$ multi-fractional Standard Model actions, which can be found in \cite{frc13}. In sections \ref{sec3}--\ref{sec5}, we calculate the bounds from, respectively, the kaon-antikaon transitions, the muon lifetime and the tau lifetime. Summary plots and conclusions are in section \ref{concl}.

%%%%%%%%%%%%%%%%%%%%%%%%%%%%%%%%%%%%%%%%%%%%%%%%%%%%%%%%%%%%%%%%%%%%%%%%%%%%%%%%%%%%%%%%%
%%%%%%%%%%%%%%%%%%%%%%%%%%%%%%%%%%%%%%%%%%%%%%%%%%%%%%%%%%%%%%%%%%%%%%%%%%%%%%%%%%%%%%%%%

\section{Theories with multi-scale spacetimes}\label{sec2}
\noindent
When one allows the dimension of spacetime to vary with the probed scale, there are two possible modifications to the effective-continuum action depending on what varies, the Hausdorff dimension $\dh$ (the scaling exponent of $D$-volumes, areas, and so on, with respect to the linear size $\ell$) or the spectral dimension $\ds$ (the energy-momentum scaling of dispersion relations), or both. In quantum gravity, usually $\ds$ is scale-dependent and $\dh$ is constant, although in multi-fractional theories and in discrete-pregeometry approaches such as group field theory, spin foams and loop quantum gravity also $\dh$ runs. 

%%%%%%%%%%%%%%%%%%%%%%%%%%%%%%%%%%%%%%%%%%%%%%%%%%%%%%%%%%%%%%%%%%%%%%%%%%%%%%%%%%%%%%%%%

\subsection{Spacetime measure}\label{smea}
\noindent
On very general and dynamics-independent grounds, when $\dh$ varies and spacetime admits a continuum limit, one can write an action such that the measure becomes $\rmd^Dx\to\rmd^Dx\,v(x)$, and the measure weight is uniquely defined parametrically as \cite{revmu,first}
\ba
v(x) &=& \prod_{\mu=0}^{D-1}v_\mu(x^\mu)\,,\qquad v_\mu(x^\mu) = \left[1+\left|\frac{x^\mu}{\ell_*^\mu}\right|^{\a_\mu-1}\right]F_\om(x^\mu)\,,\label{v}\\
F_\om(x) &=& A_0+\sum_{n=1}^{+\infty}\left[A_n\cos\left(n\om\ln\left|\frac{x}{\ell_\infty}\right|\right)+B_n\sin\left(n\om\ln\left|\frac{x}{\ell_\infty}\right|\right)\right]\,,
\ea
where $\a_\mu$ are $D$ constants such that $0<\a_\mu<1$ for all $\mu=0,1,\dots,D-1$. The values $\a_\mu=1/2$ and $\a_\mu=1/3$ are selected by quantum-gravity arguments \cite{revmu,ACCR,CaRo2}. Measures factorizable in the coordinates characterize \emph{multi-fractional spacetimes}, while the general category of \emph{multi-scale} spacetimes also includes non-factorizable versions of $v$. Effective spacetimes arising in quantum gravities are multi-scale. Many of them preserve Lorentz invariance fully or in part, and their effective measure is not factorized in all coordinates, although the measure scaling is similar or identical to \Eq{v}.

The log-oscillatory part in \Eq{v} stems from a fundamental discrete scaling symmetry $x^\mu\to \exp(-2\pi n/\om)x^\mu$ in the ultraviolet (UV). This symmetry \cite{first,cmplx} is part of the definition of multi-fractional spacetime and represents a geometry classified as a deterministic fractal. In other quantum gravity models, it is either absent or under search \cite{cmplx,StTh}. At scales $\ell_\infty\ll\ell\lesssim\ell_*$, $F_\om$ is averaged out and only its zero mode survives,
\be\label{a0}
\langle F_\om\rangle=A_0\,. 
\ee
According to the $n$-dependence of the amplitudes $A_n$ and $B_n$ and to the value of $A_0$ (1 or 0), the measure behaves in two very different ways, which have been called (respectively) deterministic picture and stochastic picture \cite{ACCR,CaRo2}. 

The \emph{deterministic picture} is realized when $A_0=1$ and $A_n$ and $B_n$ are specific functions of $n$ (usually, they decay as power laws or as exponentials). Without loss of generality \cite{revmu}, we consider a simplified isotropic profile under the coarse-graining approximation \Eq{a0}:
\be\label{v2}
v_\mu(x^\mu) = 1+\left|\frac{x^\mu}{\ell_*^\mu}\right|^{\a_\mu-1}\qquad \textrm{(deterministic view)}\,,
\ee
with the spatial $\a_i=\a$ and $\ell_*^i=\ell_*$ all equal. In the time direction, we denote $\ell_*^0=t_*$ the characteristic time scale at which effects of anomalous geometry become important. A problem with this picture in flat spacetime is that, since \Eq{v2} is a deterministic function of the coordinate $x^\mu$, the frame origin is fixed for all observers ---in \Eq{v} and \Eq{v2}, the origin is at $x^\mu=0$. As we will see below, the couplings and observables of the theory have a measure dependence. Therefore, the same experiment done at different places or times will not give the same output. For instance, an observable $\cO(t)$ may experience an absolute time running due to its measure dependence, from the big bang at $t=0$ until today 14 billion years later. This is similar to the old proposal by Dirac of running couplings \cite{Dirac1,Dirac2} and has a strong parallel with recent varying-coupling and varying-speed-of-light models \cite{frc8,Mag03,frc9}. However, from the perspective of quantum field theory on flat spacetime this arrangement may result unsatisfactory. On a curved background, the problem disappears because \Eq{v} or \Eq{v2} is the measure in the local inertial frame of the observer, so that the coordinate origin in the measure can be interpreted as the beginning of the experiment. 

The \emph{stochastic view} is realized when $A_0=0$ (log oscillations average to zero) and the amplitudes $A_n$ and $B_n$ contain random $n$-dependent phases \cite{CaRo2,cmplx}. Then, the log-oscillatory part is interpreted as a stochastic noise or fuzziness intrinsic to the fabric of spacetime. The profile \Eq{v2} is replaced by one where the power-law correction is the maximal possible fluctuation (the index $\mu$ is mute as before):
\be\label{v3}
v_\mu(x^\mu) = 1+\de v_\mu\,,\qquad |\de v_\mu|\leq \left|\frac{x^\mu}{\ell_*^\mu}\right|^{\a_\mu-1}\qquad \textrm{(stochastic view)}\,.
\ee

For our purpose, either view can be implemented, since observations will give an upper bound on $\ell_*$ and $t_*$. In the deterministic view, this implies that the approximation \Eq{v2} is sufficient, since particle-physics experiments at the LHC scales may be sensitive to $\ell_*$ but not to smaller scales (such as $\ell_\infty=\lp$ or others omitted in \Eq{v} \cite{revmu}) in the hierarchy of the measure. In the stochastic view, one uses \Eq{v3} instead of \Eq{v2} and the final upper bound is the same. In both cases, the sign in front of the power law will be unimportant, since we will compare the absolute value of the correction with the experimental error. There may be experiments where the upper and lower error bars are different and the sign of the correction becomes important, but this will not happen for the estimates in this paper. In the formul\ae\ below, we will write the geometry corrections as in \Eq{v2} for simplicity.

To conclude the review on the measure, we recall that the deep-UV scales $t_\infty$ and $\ell_\infty$ can be identified with the Planck scale, and that the time and length scales $t_*$ and $\ell_*$ are related to each other and define an energy scale $E_*$. In $c=1$ units \cite{revmu},
\be\label{tle}
t_*=\frac{\tp\ell_*}{\lp}\,,\qquad E_*:=\frac{\ep\lp}{\ell_*}=\frac{\ep\tp}{t_*}\,,
\ee
where $\tp\approx 5.3912\times 10^{-44}\,{\rm s}$, $\lp\approx 1.6163 \times 10^{-35}\,\textrm{m}$ and $\ep\approx 1.2209\times 10^{19}\,{\rm GeV}$.

%%%%%%%%%%%%%%%%%%%%%%%%%%%%%%%%%%%%%%%%%%%%%%%%%%%%%%%%%%%%%%%%%%%%%%%%%%%%%%%%%%%%%%%%%

\subsection{Multi-fractional theory with weighted derivatives}
\noindent
When also the spectral dimension $\ds$ is scale-dependent, modified dispersion relations arise from exotic kinetic terms $\phi\cK\phi$ in the action. The integral structure given by the measure weight $v(x)$ and the symmetries imposed on the Lagrangian determine the differential structure of the derivative operator $\cK$. One specific multi-scale scenario is the multi-fractional theory with weighted derivatives \cite{frc3,frc7,frc9,frc11,frc13,frc15}. Instead of going through yet another review on the subject \cite{revmu,IFWGP}, we summarize here the main features of the theory.

\subsubsection{Symmetries and dimensions}
\noindent
Poincar\'e and Lorentz invariance are broken explicitly by the measure weight $v(x)$, which selects a preferred frame where all physical observables should be computed. In this physical (also called fractional) frame, the measure weight has the profile \Eq{v} or its coarse-grained version \Eq{v2}.

In the fractional frame, clocks and rods measure a varying Hausdorff dimension of spacetime, which is
\be\label{dhmf}
\dh^{\rm UV}\simeq \a_0+(D-1)\a<D\,,\qquad \dh^{\rm IR}\simeq D\,,
\ee
where $0<\a_0,\a<1$. The spectral dimension is constant, $\ds=D$, implying that this spacetime is multi-scale, but not multi-fractal \cite{frc7}.

\subsubsection{Fractional frame and integer frame}
\noindent 
The dynamics of the theory with weighted derivatives is heavily affected by measure factors $v(x)$, non-standard kinetic terms and varying couplings. Fortunately, the problem can be simplified with a trick. There exists a non-physical frame \cite{revmu,trtls}, called integer frame or picture, which is connected by field and coupling redefinitions of the form
\be\label{phiga}
\phi_i\leftrightarrow\tilde\phi_i:=\sqrt{v}\,\phi_i\,,\qquad c_i(x)\leftrightarrow \tilde c_i:=\frac{c_i(x)}{\sqrt{v}}={\rm const}\,,
\ee
so that the fractional Standard Model reduces exactly to the ordinary one in this frame, with standard kinetic terms and constant linear gauge couplings\footnote{The theory in the integer frame is not trivialized in the presence of gravity, and there is a non-trivial coupling between matter and the non-metric structure of the geometry (i.e., the measure weight $v$) also in the integer frame \cite{revmu}.} $\tilde c_i$. The $\tilde c_i$ are constant because, by construction, the fractional-frame couplings $c_i(x)$ must carry a $\sqrt{v}$ dependence for gauge derivatives to scale homogeneously \cite{frc13}.

In the integer frame, one can use ordinary quantum-field-theory techniques and borrow all calculations from the ordinary Standard Model, until almost the end. Unlike most scalar-tensor models of gravity, where the Jordan and the Einstein frame are equivalent, there is no physical equivalence between the fractional and the integer frame and, when comparing the theory with experiments, one must revert back to the fractional frame where physical couplings are spacetime-dependent. The reason of the inequivalence is that only in the fractional frame is spacetime multi-scale and thus dimensional flow, a definitory feature of the theory, occurs. When a spacetime is endowed with one or more intrinsic scales and Lorentz invariance is broken, a special frame is selected.

\subsubsection{Observables}\label{obse}
\noindent 
Let $\ell_{\rm exp}$ and $t_{\rm exp}$ be the physical length and time scale of the physical process observed in the experiment, which are typically larger than the fundamental scales $\ell_*$ and $t_*$ of the geometry. Similarly to \Eq{tle}, all experimental scales are related to one another:
\be\label{tle2}
t_{\rm exp}=\frac{\tp\ell_{\rm exp}}{\lp}\,,\qquad E_{\rm exp}:=\frac{\ep\lp}{\ell_{\rm exp}}=\frac{\ep\tp}{t_{\rm exp}}\,,
\ee
in $c=1$ units. Therefore, for definiteness we will concentrate on $\ell_{\rm exp}$.

Suppose we want to measure an observable $\cO(c_i)$ built from some couplings $c_i$. These couplings are spacetime-dependent in the physical frame and it is very difficult to handle perturbative quantum field theory therein \cite{frc9}. By field redefinitions, one moves to the integer frame where, in the absence of gravity, the multi-fractional Standard Model reduces to the usual one with constant couplings $\tilde c_i$, indicated by a tilde. Thus, one can do all calculations in the non-physical (integer) frame to get $\cO(\tilde c_i)$ and then revert back to the physical (fractional) frame via \Eq{phiga}, to get $\cO[c_i(x)/\sqrt{v(x)}]$. Expanding powers of the measure weight \Eq{v2} as
\be
v^n(\ell_{\rm exp})=\left[1+\de v\!\left(\frac{\ell_*}{\ell_{\rm exp}}\right)\right]^n\simeq 1+ n\,\de v\!\left(\frac{\ell_*}{\ell_{\rm exp}}\right)\,,
\ee
we get
\be\nonumber
\cO(\tilde c_i)=\cO\left[\frac{c_i(\ell_{\rm exp})}{\sqrt{v(\ell_{\rm exp})}}\right]\simeq \cO[c_i(\ell_{\rm exp})]+\de\cO\left[\de v\!\left(\frac{\ell_*}{\ell_{\rm exp}}\right)\right],
\ee
where $\de\cO$ is a $\de v$-dependent correction to the standard expression. As we said above, in the stochastic picture $\de v$ is a noise with a certain scaling and it cannot be eliminated in any measurement. Therefore, while $\cO[c_i(\ell_{\rm exp})]$ is measured by the apparatus, $\de\cO[\de v(\ell_*/\ell_{\rm exp})]$ adds to the experimental noise $\de\cO_{\rm exp}$. Detection of anomalous-geometry effects would happen if $|\de\cO|>\de\cO_{\rm exp}$. If nothing unusual is observed, then the opposite inequality
\be\label{oo}
\left|\de\cO\left[\de v\left(\frac{\ell_*}{\ell_{\rm exp}}\right)\right]\right|<\de\cO_{\rm exp}
\ee
constrains the parameter space $(\a,\ell_*)$. %In this paper, we will focus on observables $\cO(G_{\rm F})$ related to weak interactions.

Typically, $\cO=av^n$, where $a$ is a constant of either sign, so that, in $D=4$ topological dimensions, to leading order in geometry corrections and using \Eq{tle} and \Eq{tle2}, from \Eq{v2} one has
\be
\de\cO\simeq b\left(\frac{t_*^{1-\a_0}}{t_{\rm exp}^{1-\a_0}}+3\frac{\ell_*^{1-\a}}{\ell_{\rm exp}^{1-\a}}\right)\cO=b\left(\frac{\ell_*^{1-\a_0}}{\ell_{\rm exp}^{1-\a_0}}+3\frac{\ell_*^{1-\a}}{\ell_{\rm exp}^{1-\a}}\right)\cO\,,
\ee
where $b=an$ is a constant of either sign. When allowing $\a_0$ to be different from $\a$, the prefactor 3 in the second term makes it dominant over the first, which can be dropped. If $\a_0=\a$, the time and spatial corrections sum together and the overall effect becomes stronger, $\de\cO\simeq 4b (\ell_*/\ell_{\rm exp})^{1-\a}\cO$. We will make the former choice, which is less restrictive. Then, the inequality \Eq{oo} becomes
\be\label{bou2}
\ell_*<\left|\frac{\de\cO_{\rm exp}}{3b\cO}\right|^{\frac{1}{1-\a}}\ell_{\rm exp}\qquad \textrm{(direct)},
\ee
while the constraints on $t_*$ and $E_*$ are derived from \Eq{bou2}:
\be\label{bou3}
t_*<\left|\frac{\de\cO_{\rm exp}}{3b\cO}\right|^{\frac{1}{1-\a}}t_{\rm exp}\,,\qquad E_*>\left|\frac{3b\cO}{\de\cO_{\rm exp}}\right|^{\frac{1}{1-\a}}E_{\rm exp}\qquad \textrm{(indirect)}.
\ee
If multi-scale effects are confined to time or space directions, a conservative one-parameter upper bound on the time scale $t_*$ and on the length scale $\ell_*$ is, respectively, \Eq{bou2} and
\be\label{bou}
t_* < \left|\frac{\de\cO}{b\cO}\right|^{\frac{1}{1-\a_0}}t_{\rm exp}\qquad \textrm{(direct)},
\ee
if we took time and spatial corrections separately. However, due to the larger numerical factor in the denominator the direct length bound \Eq{bou2} is always stronger than the indirect length bound coming from the direct time bound \Eq{bou}. Therefore, the strongest constraints come from the \emph{direct} length bound \Eq{bou2} and the \emph{indirect} time bound \Eq{bou3}.

\subsubsection{Standard Model}
\noindent 
The Standard Model of electroweak and strong interactions was developed in \cite{frc13} for this theory. All masses are constant in both frames, including the vector boson masses $M_W$ and $M_Z$, the Higgs mass and the quark masses. This is due to a conspiracy of $v$-factors from couplings and the Higgs vacuum expectation value eliding one another.

All couplings in the integer frame will be denoted with a tilde. In the specific case of the electroweak sector, the gauge couplings $g$ and $g'$ in the fractional frame are measure-dependent, while their counterparts $\tilde g=g/\sqrt{v}$ and $\tilde g'=g'/\sqrt{v}$ in the integer frame are constant. In particular, the Fermi coupling in the integer picture is \cite{frc13}
\be\label{tGF}
\tilde G_{\rm F} = \frac{\sqrt{2}\,\tilde g^2}{8M_W^2}\qquad \textrm{(integer frame, non-physical)}\,,
\ee
while the physical Fermi coupling is given by \Eq{tGF} with $\tilde g$ replaced by the physical gauge coupling $g(x)$ in the fractional frame:
\be\label{GF}
G_{\rm F}(x) = \frac{g^2(x)}{\tilde g^2} \tilde G_{\rm F}\stackrel{\textrm{\tiny\Eq{phiga}}}{=} v(x)\,\tilde G_{\rm F}\qquad \textrm{(fractional frame, physical)}\,,
\ee
where we used the fact that the $W$ mass 
\[
M_W := \frac{g' w}{2} = \frac{\tilde g'\tilde  w}{2}
\]
is constant in both frames, where $g'(x)=\sqrt{v(x)}\,\tilde g'$ is the $SU(2)_{\rm L}$ gauge coupling and $w(x)=\tilde w/\sqrt{v(x)}$ is proportional to the vacuum expectation value of the Higgs boson \cite{frc13}. In the integer frame where the theory is trivialized, the coupling $\tilde g'$ and vacuum expectation value $\tilde w$ are constant. In the physical frame, the physical quantities $g'(x)$ and $w(x)$ are spacetime-dependent, but their product is constant.

%%%%%%%%%%%%%%%%%%%%%%%%%%%%%%%%%%%%%%%%%%%%%%%%%%%%%%%%%%%%%%%%%%%%%%%%%%%%%%%%%%%%%%%%%

\subsection{Multi-fractional theory with \texorpdfstring{$q$}{}-derivatives}

\subsubsection{Symmetries and dimensions}
\noindent
The theory with $q$-derivatives \cite{revmu,frc7,frc11,frc12,frc15} is possibly the simplest of all multi-fractional proposals. It is defined by replacing the coordinates $x^\mu$ in the standard (gravitational and/or particle) action with the profiles (no summation over indices)
\be\label{geoco}
q^\mu(x^\mu)=\int^{x^\mu}\rmd {x'}^\mu\,v_\mu({x'}^\mu)\,.
\ee
All derivatives become $\p_\mu\to \p/\p q^\mu(x^\mu)=v_\mu^{-1}(x^\mu)\,\p_\mu$, hence the name of this proposal.

In this theory, the Hausdorff dimension is the same as in \Eq{dhmf}, while $\ds=\dh$ and also the spectral dimension varies with the scale \cite{frc7}. The underlying geometry is that of a multi-fractal.

\subsubsection{Fractional frame and integer frame}
\noindent
The mapping
\be\label{map}
x^\mu\leftrightarrow q^\mu(x^\mu)
\ee
is not a change of coordinates, since physical rulers and clocks are defined upon the coordinate system $x^\mu$, where spacetime is multi-scale. Contrary to this fractional frame, the integer frame spanned by the composite coordinates \Eq{geoco} is not associated with a multi-scale geometry, since all scales are hidden in the profiles $q^\mu(x^\mu)$ \cite{revmu,trtls}.

The fractional (physical) frame is defined by the coordinates $x^\mu$, while the integer frame is defined the the geometric profiles $q^\mu$, treated as scale-independent (i.e., non-composite) objects. The map \Eq{map} relates one frame to the other.

\subsubsection{Observables and Standard Model}
\noindent 
In terms of the coordinates $q^\mu$, the Standard Model is the same as the ordinary one \cite{frc12}. Therefore, one can do all calculations in the integer frame and then move back to the fractional frame, making explicit the scale dependence in the physical observables. 

In this paper, we will be interested in decay rates $\Gamma$ of particle-physics processes of the type $\Phi\to \Phi_1\Phi_2\cdots$, where $\Phi$ is a particle decaying into several products $\Phi_1$, $\Phi_2$, and so on. In the ordinary Standard Model, the inverse of the decay rate $\tilde\Gamma$ defines the lifetime $\tilde\tau$ of the particle $\Phi$, $\tilde\tau:=1/\tilde\Gamma$. However, in the theory with $q$-derivatives the time $\tilde\tau$ is not the one physically measured, which will be denoted by $\tau$ without a tilde. The two expressions are related by the mapping $q^0(t)$, which is found from (\ref{geoco}) to be
\be
\tilde\tau=q^0(\tau)\simeq \tau+\frac{t_*}{\a_0}\left(\frac{\tau}{t_*}\right)^{\a_0}\,.
\ee
Measuring $\tau$ with a $2\s$-level experimental error $2\de\tau$, no multi-scale effects are observed if
\be\label{qbou}
\frac{t_*}{\a_0}\left(\frac{\tau}{t_*}\right)^{\a_0}<2\de\tau\qquad\Rightarrow\qquad t_*<\left(\frac{2\a_0\de\tau}{\tau}\right)^{\frac{1}{1-\a_0}}\tau\,.
\ee
Bounds on the length scale $\ell_*$ and the energy scale $E_*$ can be found from the constraint \Eq{qbou} on $t_*$ via \Eq{tle}.

Note that \Eq{qbou} has a maximum at some value $\bar\a_0$ which has no significance in the theory. Therefore, for $\a_0=\bar\a_0$ one gets an absolute bound, the weakest possible constraint on the theory for a given observation. Any other value of $\a_0$ will give rise to stronger bounds.

%%%%%%%%%%%%%%%%%%%%%%%%%%%%%%%%%%%%%%%%%%%%%%%%%%%%%%%%%%%%%%%%%%%%%%%%%%%%%%%%%%%%%%%%%
%%%%%%%%%%%%%%%%%%%%%%%%%%%%%%%%%%%%%%%%%%%%%%%%%%%%%%%%%%%%%%%%%%%%%%%%%%%%%%%%%%%%%%%%%

\section{\texorpdfstring{$K^{0}-\bar{K}^{0}$}{} transitions}\label{sec3}

%%%%%%%%%%%%%%%%%%%%%%%%%%%%%%%%%%%%%%%%%%%%%%%%%%%%%%%%%%%%%%%%%%%%%%%%%%%%%%%%%%%%%%%%%

\subsection{Weighted derivatives}
\noindent
It is well known that the strangeness (S) quantum number is not conserved by weak interactions. The $K^{0}-\bar{K}^{0}$ transition violates the S-number by two units. In the integer frame, the theory is exactly the same as the ordinary Standard Model and all the formul\ae\ below can be taken from the literature (see \cite{pdg} for more details on the kaon-antikaon transitions). In particular, the $2\times 2$  mixing matrix of $K^{0}-\bar{K}^{0}$ can be written as 
\be \label{Mass}
\tilde{\mathcal{M}}_{K} = \left(\begin{array}{cc} \tilde{M}_K
 & \tilde{M}_{12}
\ \\ \tilde{M}_{12}^{*} & \tilde{M}_K \\
\end{array} \right)  - \frac{\rmi}{2}\left( \begin{array}{cc} \tilde{\Gamma} & \tilde{\Gamma}_{12} \\ \tilde{\Gamma}_{12}^{*} & \tilde{\Gamma} \end{array} \right)\,,
\ee
where the diagonal elements $\tilde{M}_K$ (kaon mass) and $\tilde{\Gamma}$ (kaon decay rates) are real parameters in the integer frame. The decay rates in the fractional frame $\Gamma,\Gamma_{12}$ are expected to be different from $\tilde{\Gamma},\tilde{\Gamma}_{12}$ with a factor we will estimate later on. On the other hand, the kaon and antikaon masses are the same in both frames: $M_K=\tilde{M}_K$. The kaon mass is generated by the quantum chromodynamics (QCD) phase dimensional transmutation and it is proportional to the QCD phase, while quark-mass contributions are negligible. Both the QCD scale and the quark masses do not change in the two frames. In the case of the QCD scale, this is because it is given by the chiral symmetry breaking condensate
\be\label{qqbar}
\langle 0| \bar{q}q |0\rangle=  \langle \tilde{0}|v^{-1}({\bf x})v({\bf x})\,\tilde{\bar{q}}\tilde{q}|\tilde{0}\rangle=\Lambda_{\rm QCD}^{3}\, , 
\ee
where the $v({\bf x})=v_1(x^1)v_2(x^2)v_3(x^3)$ factor (the spatial part of the measure weight) comes from two quark-field transformations and $v^{-1}({\bf x})$ comes from the transformation of the vacuum state $|0\rangle$ in the fractional frame to the integer one $|\tilde{0}\rangle$. However, subtly, the fact that the integer and the fractional frames have different spacetime measures does not imply that the off-diagonal masses will be the same in the two frames as well: we will see that they are generated from the electroweak mixing among quarks. This will imply that the mass eigenvalues of the mass matrix (\ref{Mass}) change in the fractional frame. 

Before going into the details of the analysis, let us make an important observation: since the $v$ factor enters democratically between diagonal terms and off-diagonal terms, but it is not factorized out democratically, there is no violation of hermiticity in the Hamiltonian mass matrix in the fractional frame. In other words, even if Lorentz symmetry is violated no CPT violating phases appear, at least at the one-loop level in the strong and electroweak sectors. In \cite{frc13}, CPT invariance was shown only at the classical level.

The physical eigenstates in the integer frame are 
\be \label{states}
|\tilde{K}_{\rm L,S}\rangle=\frac{1}{\sqrt{|\tilde p|^{2}+|\tilde q|^{2}}}\,(\tilde p|K^{0}\rangle \mp \tilde q |\bar{K}^{0}\rangle)\,, 
\ee
where 
\be \label{jkk}
\frac{\tilde{q}}{\tilde{p}}:= \frac{1-\bar{\epsilon}}{1+\epsilon}=\left( \frac{\tilde{M}_{12}^{*}-\frac{i}{2}\tilde{\Gamma}_{12}^{*}}{\tilde{M}_{12}-\frac{i}{2}\tilde{\Gamma}_{12}}\right)^{\frac12}. 
\ee
The imaginary parts of $\tilde{M}_{12}$ and $\tilde{\Gamma}_{12}$ are CP violating: if ${\rm Im}(\tilde{M}_{12})=0={\rm Im}(\tilde{\Gamma}_{12})$, then $\tilde{q}=\tilde{p}$ and the two eigenstates $\tilde{K}_{\rm L,S}$ trivialize to two orthogonal eigenvectors (a CP-odd and a CP-even state). In other words, the departure from CP-symmetry is parametrized by $\tilde{p},\tilde{q}$ or ${\bar{\epsilon}}$:
\be \label{horto}
 \langle \tilde{K}_{\rm L}| \tilde{K}_{\rm S} \rangle= \langle K_{\rm L}| K_{\rm S} \rangle=\frac{|\tilde{p}|^{2}-|\tilde{q}|^{2}}{|\tilde{p}|^{2}+|\tilde{q}|^{2}}\,.
\ee
Thus, in the theory with weighted derivatives the product of states is the same in both reference frames, since any multiplicative dependence of $q,p$ from $v(x)$ is elided in the numerator and the denominator. 

%The departure from $p/q$ unity can be measured 
%by CP-violating asymmetry in certain flavour-decays such as 
%\be \label{KKbar}
%K^{0}\rightarrow \pi^{-}l^{+}\nu_{l},\,\,\,\, \bar{K}^{0}\rightarrow \pi^{+}l^{-}\bar{\nu}_{l},
%\ee
%In the SM, 
%the decay rates of kaon and antikaon are CPT preserving and CP violating, i.e.
%\be \label{delta}
%\delta \equiv \frac{\Gamma(K_{L}\rightarrow \pi^{-}l^{+}\nu_{l})-\Gamma(K_{L}\rightarrow \pi^{+}l^{-}\bar{\nu}_{l})}{\Gamma(K_{L}\rightarrow \pi^{-}l^{+}\nu_{l})+\Gamma(K_{L}\rightarrow \pi^{+}l^{-}\bar{\nu}_{l})}=\frac{|p|^{2}-|q|^{2}}{|p|^{2}+|q|^{2}}\equiv \frac{2{\rm Re}(\bar{\epsilon})}{(1+\bar{\epsilon})^{2}}\, .
%\ee

The CP-violation signal is encoded into asymmetry parameters in hadronic channels:
\be \label{Kpipi}
\eta_{+-}=\frac{\Gamma(K_{\rm L}\rightarrow \pi^{+}\pi^{-})}{\Gamma(K_{\rm S}\rightarrow \pi^{+}\pi^{-})}=|\eta_{+-}|\rme^{\rmi\phi_{+-}}\simeq \epsilon+\epsilon',\,\,\, 
\eta_{00}=\frac{\Gamma(K_{\rm L}\rightarrow \pi^{0}\pi^{0})}{\Gamma(K_{\rm S}\rightarrow \pi^{0}\pi^{0})}=|\eta_{00}|\rme^{\rmi\phi_{00}}\simeq \epsilon-2\epsilon'\, , 
\ee
where $\epsilon$ and $\epsilon'$ are the CP violating parameters. 
%\be \label{epsilon}
%\epsilon=\bar{\epsilon}+i\zeta_{0},\,\,\,
%\epsilon'=\frac{i}{\sqrt{2}}\omega(\zeta_{2}-\zeta_{0}),\,\,\,
%\omega=\frac{{\rm Re}(\Gamma_{2})}{{\rm Re}(\Gamma_{0})}e^{i(\delta_{2}-\delta_{0})}\, . 
%\ee
%The phase of the $\epsilon$ parameter is the 
%superweak phase
%\be \label{SW}
%\phi_{SW}=\tan^{-1}2\Delta m_{K}/\Delta \Gamma\, . 
%\ee
Experimentally, ${\rm Re}(\epsilon'/\epsilon)$ is a measure of the CP violation determined from decay rates as 
\be \label{Gammam}
\frac{\Gamma(K_{\rm L}\rightarrow \pi^{+}\pi^{-})/\Gamma(K_{\rm S}\rightarrow \pi^{+}\pi^{-})}{\Gamma(K_{\rm L}\rightarrow \pi^{0}\pi^{0})/\Gamma(K_{\rm S}\rightarrow \pi^{0}\pi^{0})}=\Big|\frac{\eta_{+-}}{\eta_{00}}\Big|^{2}\simeq 1+6\,{\rm Re}\left(\frac{\epsilon'}{\epsilon}\right). 
\ee
For  small $|\epsilon'/\epsilon|$, ${\rm Im}(\epsilon'/\epsilon)$ is related to the phases of $\eta_{+-}$ and $\eta_{00}$ by 
\be \label{phipmzz}
\phi_{+-}\simeq \phi_{\epsilon}+{\rm Im}\left(\frac{\epsilon'}{\epsilon}\right), \qquad \phi_{00}\simeq \phi_{\epsilon}-2\,{\rm Im}\left(\frac{\epsilon'}{\epsilon}\right),\qquad \phi_{00}-\phi_{+-}\simeq -3\,{\rm Im}\left(\frac{\epsilon'}{\epsilon}\right). 
\ee

In the multi-fractional Standard Model, observable quantities depend on decay rates in the fractional reference frame. The dependence of kaon parameters from $v(x)$ is determined by the microscopic parameters of the theory. 
\begin{figure}[t]
\centerline{ \includegraphics [height=3.5 cm,width=0.8\columnwidth]{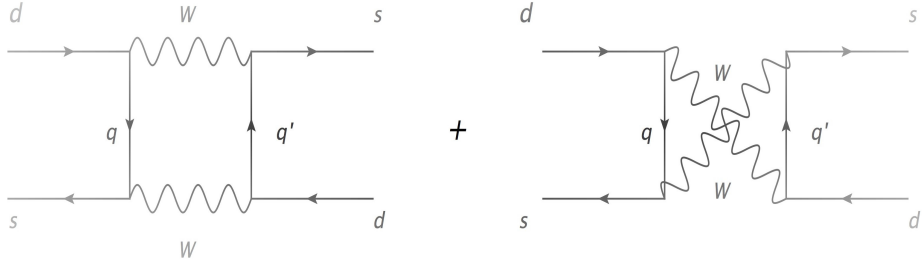}} %.pdf
\vspace*{-1ex}
\caption{\label{fig1} $K^{0}-\bar{K}^{0}$ transitions. $q,q'=\{u,c,t\}$ are the quark species contributing to the virtual box diagram. The dominant channel is given by the exchange of charm quarks.}
\end{figure}

The Standard Model diagrams contributing to the $K^{0}-\bar{K}^{0}$ transitions are displayed in Fig.\ \ref{fig1}, which corresponds to the effective Hamiltonian interaction terms
\be \label{HS}
H_{\Delta {\rm S}=2}=\sum_{q,q'}\frac{\tilde G_{\rm F}^{2}}{4\pi^{2}}|V_{qd}V^{*}_{qs}V_{q'd}V^{*}_{q's}  |m_{q}m_{q'} [\bar{d}\gamma^{\mu}(1+\gamma_5)s][\bar{d}\gamma_{\mu}(1+\gamma_5)s]\,, 
\ee
where $\tilde G_{\rm F}$ is the Fermi coupling (a constant, in this case), $V_{qq'}$ are the Cabibbo--Kobayashi--Maskawa (CKM) matrix elements of two quark species $q,q'$, and $m_{q,q'}$ are the quark masses. Let us remark that, in (\ref{HS}), the quark masses and CKM matrix elements are invariant under the frame change from integer to fractional and vice versa. Also, the Hamiltonian interaction violates the Strange number by two units, as well as CP symmetry. The effective Hamiltonian must be evaluated as a matrix element $\langle K^{0}|H_{\Delta {\rm S}=2}|\bar{K}^{0}\rangle$. Using the so-called {\it vacuum saturation approximation} (see \cite{Mohapatra:1968zz} for details), one obtains
\be \label{KKKK}
\Delta M_{K}\simeq M_{K_{\rm L}}-M_{K_{\rm S}}\simeq \sum_{q,q'}\frac{\tilde G_{\rm F}^{2}}{6\pi^{2}}f_{K}^{2}|V_{qd}V^{*}_{qs}V_{q'd}V^{*}_{q's}  |M_K m_{q}m_{q'}\,,
\ee
where $f_{K}$ and $M_K$ are, respectively, the kaon decay constant and mass. The dominant contribution is provided by the charm and the top virtual quarks.

From a more detailed one-loop calculation, one obtains 
\be\label{Kkbar}
\Delta M_{K}=\frac{\tilde G_{\rm F}^{2}}{6\pi^{2}}M_{W}^{2}M_{K}f_{K}^{2}B\,F(x_{i},\theta_{j}),
\ee
where
\ba
F(x_{i},\theta_{j})&=&[({\rm Re}\lambda_{c})^{2}+({\rm Im}\lambda_{c})^{2}]\eta_{1}f(x_{c})+[({\rm Re}\lambda_{t})^{2}-({\rm Im}\lambda_{t})^{2}]\eta_{2}f(x_{t})\nonumber\\
&&+2({\rm Re}\lambda_{c}{\rm Re}\lambda_{t}-{\rm Im}\lambda_{c}{\rm Im}\lambda_{t})\eta_{3}f(x_{t},x_{c})\,,\label{xtheta}\\
f(x_{i}) &=& x_{i}\Big[\frac{1}{4}+\frac{9}{4}(1-x_{i})^{-1}-\frac{3}{2}(1-x_{i})^{-2}\Big]+\frac{3}{2}\Big[\frac{x_{i}}{x_{i}-1}\Big]^{3}\ln x_{i}\,,\label{fx}\\
f(x_{i},x_{j})&=& x_{i}x_{j}\Big\{\Big[\frac{1}{4}+\frac{3}{2}(1-x_{i})^{-1}-\frac{3}{4}(1-x_{i})^{-2}\Big] \frac{\ln x_{j}}{x_{j}-x_{i}}+(x_{j}\leftrightarrow x_{i})\nonumber\\
&&-\frac{3}{4}[(1-x_{j})(1-x_{i})]^{-1}\Big\},\label{fxx}
\ea
$\lambda_{i}=V_{id}^{*}V_{is}$, $x_{i}=m_{q_{i}}^{2}/M_{W}^{2}$, $\eta_{i}$ are hadronic QCD corrections  ($\eta_{1}\simeq 0.85$, $\eta\simeq 0.6$ and $\eta_{3}\simeq 0.39$), and $B$ is a
scale- and scheme-independent quantity \cite{DGH,Gam03} that captures numerical factors of the matrix element $\langle K^{0}|\bar{d}\gamma^{\mu}(1+\gamma_{5})s\bar{d}\gamma_{\mu}(1+\gamma_{5})s|\bar{K}^{0}\rangle$.

On the other hand, the kaon decay into the hadronic channel $K\rightarrow \pi\pi$ is 
\be \label{Kpipi2}
H_{\rm eff}(\Delta {\rm S}=1)=\frac{\tilde{G}_{\rm F}}{\sqrt{2}}V_{ud}V^{*}_{us}\sum_{i=1}^{10}C_{i}(\mu)Q_{i}(\mu)+{\rm h.c.}\,,
\ee
where $C_{i}$ are the Wilson coefficient, $Q_{i}$ are the local four-fermion operators controlling the decay and ``h.c.'' denotes hermitian conjugate.

%Up to here, we have reviewed the computation in the ordinary Standard Model.
In the multi-fractional theory with weighted derivatives, all the aforementioned calculations are valid in the integer frame, but \Eq{Kkbar} is not what is observed in experiments. To extract the physical observable, we must reconsider carefully all the relations we wrote down explicitly and keep track of their measure dependence. It turns out that the only non-trivial modification comes from the relation \Eq{GF} between $\tilde G_{\rm F}$ and the physical Fermi coupling $G_{\rm F}(x)$. The Wilson coefficients multiplying the operators in \Eq{Kpipi2} do not depend on $v(x)$ at the leading order, i.e., they can be considered as constant factors. This implies 
\be \label{Kpip}
\tilde{\Gamma}(K\rightarrow \pi\pi)=\frac{1}{v^{2}(x)}\Gamma(K\rightarrow \pi\pi)=:\frac{1}{v^{2}(x)}\frac{1}{\tau_{\rm S}}\simeq \left(1-2\frac{t_*^{1-\a_0}}{t^{1-\a_0}}-6\frac{\ell_*^{1-\a}}{\ell^{1-\a}}\right)\frac{1}{\tau_{\rm S}}\,.
\ee
The ``$-$'' signs in the multi-fractional corrections hold in the deterministic view, while they become a ``$\pm$'' in the stochastic view (see section \ref{smea}). Since we will take the absolute value, this difference is irrelevant.

As we stressed above, the masses of the $W$ boson and of the quarks are constant in both frames \cite{frc13}, and so is the kaon mass $M_K$. Regarding the mass-splitting formula \Eq{KKKK}, in the integer frame (ordinary Standard Model) $f_{K}$ is defined as the correlator $\langle 0|j_\mu|K\rangle= f_K q_\mu$, where $|0\rangle$ is the leptonic state. A similar expression holds for the pion. Therefore, at the tree level $f_K$ depends only on the properties of the meson and scales as its mass. Thus $f_K$ remains constant also in the fractional frame, and the only effect of geometry comes from the Fermi coupling:
\be\label{Kkbar2}
\Delta M_{K}=\frac{1}{v^2(x)}\frac{G_{\rm F}^{2}}{6\pi^{2}}M_{W}^{2}M_{K}f_{K}^{2}\frac{m_{c}^{2}}{M_{W}^{2}}c_{c}^{2}s_{c}^{2}
\simeq\left(1-2\frac{t_*^{1-\a_0}}{t^{1-\a_0}}-6\frac{\ell_*^{1-\a}}{\ell^{1-\a}}\right)\Delta M_{K}^0\,,
\ee
where $\Delta M_{K}^0$ is the mass splitting of the standard theory with physical Fermi coupling. 

%{\bf How we put the bound: using limits from kaon decays in leptonic channels
%we can measure $\bar{\epsilon}$. From it, we can extract the limit on the 
%$\Delta m_{K}$ by using $K$-decays in hadronic channels (pions). }

%\be \label{epsilon}
%\bar{\epsilon}\simeq e^{i\phi_{SW}}\frac{{\rm Im}(M_{12})-\frac{i}{2}{\rm Im}(\Gamma_{12})}{\sqrt{\Delta M^{2}+\frac{1}{4}\Delta \Gamma^{2}}},\,\,\,
%\phi_{SW}={\rm arctan}\left( \frac{-2\Delta M}{\Delta \Gamma}\right)\, . 
%\ee

We can put stringent constraints on the multi-fractional scales from KTeV data \cite{Abouzaid:2010ny,pdg}, measuring the decay rates of 
kaons into two pions. The best fit of KTeV with $\tau_{\rm S},\Delta, \phi_{\epsilon}, \epsilon'/\epsilon$ parameters is as follow (in $\hbar=1$ units):
\ba
\tau_{\rm S}&=&(89.598\pm 0.070)\times 10^{-12}\,{\rm s}\quad \Rightarrow\quad \frac{\de\tau_{\rm S}}{\tau_{\rm S}}\approx 8\times 10^{-4}\,,\label{tauS}\\
\Delta M_{K}&=&(5279.7\pm 19.5)\times 10^{6}\,{\rm s}^{-1}\quad \Rightarrow\quad \frac{\de\Delta M_{K}}{\Delta M_{K}}\approx 4\times 10^{-3}\,,\label{Deltam}%\\
%\phi_{\epsilon}&=&(43.86\pm 0.63)^{\circ}\,,\label{phie}\\
%{\rm Re}(\epsilon'/\epsilon)&=&(21.10\pm 3.43)\times 10^{-4}\,,\label{Reee}\\
%{\rm Im}(\epsilon'/\epsilon)&=&(-17.20\pm 20.20)\times 10^{-4}\,, \label{Im}
\ea
$\phi_{\epsilon}=(43.86\pm 0.63)^{\circ}$, ${\rm Re}(\epsilon'/\epsilon)=(21.10\pm 3.43)\times 10^{-4}$, ${\rm Im}(\epsilon'/\epsilon)=(-17.20\pm 20.20)\times 10^{-4}$. The fit is performed using the oscillating function counting for the number of pion events: 
\ba
N^{\pi\pi}(p,t)&=& N_{0}\mathcal{F}(p)T_{\rm reg}(p)\times [|\rho(p)|^{2}\exp(-t/\tau_{\rm S})+|\eta|^{2}\exp(-t/\tau_{\rm L})\nonumber\\
&&+2|\rho(p)||\eta| \cos[\Delta M_{K} t +\phi_{p}(p)-\phi_{\eta}(p)]\,{\rm exp}(-t/\tau)\,,\nonumber%\label{oscillation}
\ea
where $p$ is the kaon momentum, $N_{0}$ is a normalization factor, $\rho(p)$ is the momentum-dependent coherent regeneration amplitude in the experiment, $\phi_{\rho}(p)={\rm arg}(\rho)$, $1/\tau=(1/\tau_{\rm S}+1/\tau_{\rm L})/2$, $T_{\rm reg}(p)$ is the relative kaon flux transmission in the regenerator beam, and $\mathcal{F}(p)$ is the kaon flux function. The time parameter here is  $t=z\, M_{K}/p$, where $z$ is the interaction vertex position in the detector.

In the multi-fractional model with weighted derivatives, $\phi_{\epsilon}$ and $\epsilon'/\epsilon$ are dimensionless quantities independent of the measure weight, which is the reason why we did not decorate them with a tilde. This implies that the CP violating phases are not spacetime dependent in the fractional frame and do not give constraints on new physics. However, $\tau_{\rm S}$ and $\Delta M_{K}$ are both modified as in (\ref{Kpip}) and (\ref{Kkbar2}), respectively. Bounds on $v(x)$ can be inferred from the best fit of the observables $\tau_{\rm S}$ and $\Delta M_{K}$. The relative error on the mass splitting \Eq{Deltam} is slightly bigger than for $\tau_{\rm S}$ and it gives milder bounds. Therefore, we will use $\tau_{\rm S}$. From \Eq{Kpip} and dropping the time correction as argued in section \ref{obse}, we find 
\be
t_*<\left(\frac{\de \tau_{\rm S}}{3\tau_{\rm S}}\right)^{\frac{1}{1-\a}}t_{K\bar K},\qquad \ell_*<\left(\frac{\de \tau_{\rm S}}{3\tau_{\rm S}}\right)^{\frac{1}{1-\a}}\ell_{K\bar K}\,,\qquad E_*=\frac{\ep\lp}{\ell_*}>\left(\frac{\de \tau_{\rm S}}{3\tau_{\rm S}}\right)^{\frac{1}{1-\a}}E_{K\bar K}\,,
\ee
where we used the experimental error in \Eq{tauS} at the $2\s$-level (hence the factor $2/6=1/3$). Taking the lifetime $\tau_{\rm S}$ or the mass splitting $\Delta M_{K}^{-1}$ as the characteristic time scale $t_{K\bar K}$ leads to very weak bounds, while taking instead the mass of the kaon,
\be
E_{K\bar K}=M_K\approx 494\,{\rm MeV}\,,\qquad t_{K\bar K}=\frac{\tp\ep}{E_{K\bar K}}\approx 10^{-24}\,{\rm s}\,,\qquad \ell_{K\bar K}=\frac{\lp\ep}{E_{K\bar K}}\approx 4\times 10^{-16}\,{\rm m}\,,
\ee
we obtain the bounds 
\ba
\hspace{-1.3cm}&&t_* < 3\times 10^{-28}\,{\rm s}\,,\qquad \ell_*<10^{-19}\,{\rm m}\,,\qquad E_*> 1.9\,{\rm TeV}\,,\qquad \a_0,\a\ll\frac12\,,\label{tEs}\\
\hspace{-1.3cm}&&t_* < 9\times 10^{-32}\,{\rm s}\,,\qquad \ell_*<3\times 10^{-23}\,{\rm m}\,,\qquad E_*>7\times 10^{6}\,{\rm GeV}\,,\qquad \a_0=\frac12=\a\,,\label{c3}
\ea
which are stronger than previous constraints from any other observation \cite{revmu} (except the $\a_0=1/2$ bounds from the fine-structure constant; see section \ref{concl}) but weaker than the constraints from the muon lifetime we will find below.

\subsection{\texorpdfstring{$q$}{}-derivatives}
\noindent 
In the theory with $q$-derivatives, there is no correction from couplings, which are constant both in the integer and in the fractional frame. However, particle lifetimes are measured differently in the two frames and, after replacing $\tau_{\rm S}$ above (it should wear a tilde) with $q^0(\tau_{\rm S})$, we get the bound \Eq{qbou}. For $\a_0=\bar\a_0\approx 0.11$, we get an absolute bound for the theory:
\be
t_* < 5\times 10^{-15}\,{\rm s}\,,\qquad \ell_*<2\times 10^{-6}\,{\rm m}\,,\qquad E_*>10^{-10}\,{\rm GeV}\,,
\ee
while for $\a_0=1/2$
\be
t_* < 5\times 10^{-17}\,{\rm s}\,,\qquad \ell_*<2\times 10^{-8}\,{\rm m}\,,\qquad E_*> 10^{-8}\,{\rm GeV}\,,\qquad \a_0=\frac12\,.
\ee

%%%%%%%%%%%%%%%%%%%%%%%%%%%%%%%%%%%%%%%%%%%%%%%%%%%%%%%%%%%%%%%%%%%%%%%%%%%%%%%%%%%%%%%%%
%%%%%%%%%%%%%%%%%%%%%%%%%%%%%%%%%%%%%%%%%%%%%%%%%%%%%%%%%%%%%%%%%%%%%%%%%%%%%%%%%%%%%%%%%

\section{Muon lifetime}\label{sec4}

%%%%%%%%%%%%%%%%%%%%%%%%%%%%%%%%%%%%%%%%%%%%%%%%%%%%%%%%%%%%%%%%%%%%%%%%%%%%%%%%%%%%%%%%%

\subsection{Weighted derivatives}
\noindent 
In the ordinary Standard Model, the muon decay rate $\Gamma$ is calculated from the $W$-mediated decay process $\mu^- \rightarrow e^- \bar{\nu}_e \nu_{\mu}$. Neglecting the masses of the electron $e^-$ and the neutrino $\nu_e$, one has
\be
\Gamma = \frac{\tilde G_{\rm F}^2 m_{\rm mu}^5}{192\pi^3}+\textrm{radiative corrections}\,,
\ee
where $\tilde G_{\rm F}$ is constant. In the multi-fractional theory with weighted derivatives, this expression is valid in the integer frame, where all couplings are constant. However, this frame is non-physical and to get the expression to be compared with experiments we have to use \Eq{GF} to transform back to the fractional (physical) frame:
\be
\Gamma = \frac{1}{v^2(x)}\frac{G_{\rm F}^2 m_{\rm mu}^5}{192\pi^3}+\textrm{radiative corrections}\,,
\ee
where $G_{\rm F}$ is the bare Fermi coupling. The muon mass $m_{\rm mu}$ is constant in both frames. Once again, it is worth to emphasize that $1/\Gamma$ is the physical mean lifetime of the muon, while $G_{\rm F}(x)$ and $m_{\rm mu}$ are the observables measured in experiments. In particular, the non-constant Fermi coupling $G_{\rm F}$ depends on the spacetime scale at which one is taking measurements. The factor $1/v^2(x)=1+({\rm corrections})$ gives a contribution that cannot be greater than the experimental error. From here, we can place a bound on the parameters $\a_0$, $\a$, $t_*$ and $\ell_*$ in $v$.

Using \Eq{v2}, the mean lifetime of the muon is (in $\hbar=1$ units)
\be
\tau_{\rm mu}:=\frac{1}{\Gamma}=\frac{192\pi^3}{G_{\rm F}^2 m_{\rm mu}^5}v^{2}(x)
= \frac{192\pi^3}{G_{\rm F}^2m_{\rm mu}^5}\left(1+\left|\frac{t}{t_*}\right|^{\a_0-1}\right)^{2}\left[\prod_{i=1}^3 \left(1+\left|\frac{x^i}{\ell_*}\right|^{\a-1}\right)\right]^{2}\,,\label{restim}
\ee
both in ordinary Minkowski and in multi-fractional spacetime with weighted derivatives. Experiments yield an estimate of the Fermi constant $G_{\rm F}= 1.1663787(6)\times 10^{-5}\,\textrm{GeV}^{-2}$ and of the muon mass $m_{\rm mu} = 105.6583745(24) \,\textrm{MeV}$, giving a $\tau_{\rm mu}\approx\tau_0=2.1969811(22)\times 10^{-6}\,{\rm s}$ with an error $\delta\tau_{\rm mu}\approx 2.2\times 10^{-12}\,{\rm s}$ at the $1\s$-level \cite{pdg}.

We can place constraints on the theory by considering the $2\s$-level experimental relative error $2\de\tau_{\rm mu}/\tau_{\rm mu}\approx 2\times 10^{-6}$ on the muon lifetime as an upper bound on multi-scale-geometry effects:
\be
2\de\tau_{\rm mu} \gtrsim \frac{192\pi^3\tp\ep}{G_{\rm F}^2m_{\rm mu}^5}\left(2\frac{t_*^{1-\a_0}}{t_{\rm mu}^{1-\a_0}}+6\frac{\ell_*^{1-\a}}{\ell_{\rm mu}^{1-\a}}\right)=: \tau_{\rm mu}\left(2\frac{t_*^{1-\a_0}}{t_{\rm mu}^{1-\a_0}}+6\frac{\ell_*^{1-\a}}{\ell_{\rm mu}^{1-\a}}\right),
\ee
where we inserted appropriate Planck units to make the constants in brackets dimensionless, and $\tau_{\rm mu}\simeq \tau_0$ is the standard Standard-Model muon lifetime. The one-parameter upper bounds on the geometry scales are
\be\label{mubou}
t_*< \left(\frac{\de\tau_{\rm mu}}{3\tau_{\rm mu}}\right)^{\frac{1}{1-\a}}t_{\rm mu}\,,\qquad \ell_* < \left(\frac{\de\tau_{\rm mu}}{3\tau_{\rm mu}}\right)^{\frac{1}{1-\a}}\ell_{\rm mu}\,,\qquad E_*> \left(\frac{3\tau_{\rm mu}}{\de\tau_{\rm mu}}\right)^{\frac{1}{1-\a}}E_{\rm mu}\,.
\ee

The scales $t_{\rm mu}$ and $\ell_{\rm mu}$ are those typical of muonic processes. If we set $t_{\rm mu}=\tau_0$, we would have $t_{\rm mu}=\tau_0\approx 10^{-6}\,{\rm s}$ and $\ell_{\rm mu}\approx 300\,{\rm m}$. As functions of the fractional exponents, the bounds \Eq{mubou} would be much weaker than if we set the energy scale to the mass of the muon, $E_{\rm mu}=m_{\rm mu}$. In the latter case,
\be
t_{\rm mu}=\frac{\tp\ep}{m_{\rm mu}}\approx 6\times 10^{-24}\,{\rm s}\,,\qquad \ell_{\rm mu}=\frac{\lp\ep}{m_{\rm mu}} \approx 2\times 10^{-15}\,{\rm m}.
\ee
Then, from \Eq{mubou} we get
\be\label{c1}
t_* < 2\times 10^{-30}\,{\rm s}\,,\qquad \ell_*< 6\times 10^{-22}\,{\rm m}\,,\qquad E_*>3\times 10^5\,{\rm GeV}\,,\qquad \a_0,\a\ll\frac12\,,
\ee
where the bounds on $t_*$ and $E_*$ are indirect and the one on $\ell_*$ is direct. These correspond to scales above the LHC center-of-mass energy and are the strongest absolute ($\a$-independent) bounds to date for the multi-fractional theory with weighted derivatives.

On the other hand, for the intermediate value $\a_0=1/2=\a$, we get (here the bound on $t_*$ is derived from that on $\ell_*$)
\be\label{c2}
t_* < 7\times 10^{-37}\,{\rm s}\,,\qquad \ell_*<2\times 10^{-28}\,{\rm m}\,,\qquad E_*>9 \times 10^{11}\,{\rm GeV}\,,\qquad \a_0=\frac12=\a\,,
\ee
slightly below the grand-unification scale.

Note that both \Eq{c1} and \Eq{c2} are considerably stronger than their counterparts in the theory with $q$-derivatives for the same observation \cite{frc13}. Other experiments give stronger bounds for the theory with $q$-derivatives, but not for the theory with weighted derivatives \cite{revmu}.

We can also get an upper bound on the Hausdorff dimension of spacetime $\dh^{\rm UV}$ in the ultraviolet by inverting \Eq{bou} and \Eq{bou2} and assuming the smallest admissible $\ell_*$ to be the Planck scale, $\ell_*=\lp$:
\be
\a_0<\frac{\ln\frac{2\tp}{\de\tau}}{\ln\frac{\tp}{\tau_0}}\approx 0.84\,,\qquad\a<\frac{\ln\left(\frac{\lp}{\ell_{\rm mu}}\frac{6\tau_0}{\de\tau}\right)}{\ln\frac{\lp}{\ell_{\rm mu}}}\approx 0.68\,,\qquad \dh^{\rm UV}=\a_0+3\a<2.87\,,\label{bou23}
\ee
if $0<\a_0,\a<1$. Note that both bounds are direct, hence the one in the time direction is weaker. As in \cite{frc14}, as soon as one allows the spacetime dimension to vary, one can obtain counter-intuitive \emph{upper} bounds on the dimension in the UV.

%%%%%%%%%%%%%%%%%%%%%%%%%%%%%%%%%%%%%%%%%%%%%%%%%%%%%%%%%%%%%%%%%%%%%%%%%%%%%%%%%%%%%%%%%

\subsection{\texorpdfstring{$q$}{}-derivatives}
\noindent 
Applying \Eq{qbou} to $\tau=\tau_{\rm mu}$ and $\de\tau_{\rm mu}$, the loosest bound is obtained for $\bar\a_0\approx 0.06$:
\be
t_* < 10^{-13}\,{\rm s}\,,\qquad \ell_*<3\times 10^{-5}\,{\rm m}\,,\qquad E_*>7 \times 10^{-12}\,{\rm GeV}\,,
\ee
while for $\a_0=1/2$
\be
t_* < 2\times 10^{-18}\,{\rm s}\,,\qquad \ell_*<7\times 10^{-10}\,{\rm m}\,,\qquad E_*>3 \times 10^{-7}\,{\rm GeV}\,,\qquad \a_0=\frac12\,.
\ee
These bounds \cite{frc12,frc13} are much weaker than those in the theory with weighted derivative.

%%%%%%%%%%%%%%%%%%%%%%%%%%%%%%%%%%%%%%%%%%%%%%%%%%%%%%%%%%%%%%%%%%%%%%%%%%%%%%%%%%%%%%%%%
%%%%%%%%%%%%%%%%%%%%%%%%%%%%%%%%%%%%%%%%%%%%%%%%%%%%%%%%%%%%%%%%%%%%%%%%%%%%%%%%%%%%%%%%%

\section{Tau lifetime}\label{sec5}

%%%%%%%%%%%%%%%%%%%%%%%%%%%%%%%%%%%%%%%%%%%%%%%%%%%%%%%%%%%%%%%%%%%%%%%%%%%%%%%%%%%%%%%%%

\subsection{Weighted derivatives}
\noindent 
The tau decay rate $\Gamma_{\rm tau}$ can be calculated from processes such as the electron channel $\tau^- \rightarrow \nu_\tau e^-\bar{\nu}_e$ and the muon channel $\tau^- \rightarrow \nu_\tau \mu^-\bar{\nu}_\mu$. In the ordinary Standard Model, $\Gamma_{\rm tau}$ is exactly the same as the muon decay rate with the muon mass replaced by the tau mass $m_{\rm tau}$. Therefore, the bounds have the same functional expression as in \Eq{mubou}:
\be\label{taubou}
t_*< \left(\frac{\de\tau_{\rm tau}}{3\tau_{\rm tau}}\right)^{\frac{1}{1-\a}}t_{\rm tau}\,,\qquad \ell_* < \left(\frac{\de\tau_{\rm tau}}{3\tau_{\rm tau}}\right)^{\frac{1}{1-\a}}\ell_{\rm tau}\,,\qquad E_*> \left(\frac{3\tau_{\rm tau}}{\de\tau_{\rm tau}}\right)^{\frac{1}{1-\a}}E_{\rm tau}\,.
\ee
In \Eq{taubou}, we use the value $\tau_{\rm tau}=(2.903\pm0.005)\times10^{-15}\,{\rm s}$ taken from \cite{pdg}, while $2\de\tau_{\rm tau}=10^{-17}\,{\rm s}$ denotes the $2\s$-level error. Furthermore, we take the tau mass $m_{\rm tau}\approx 1.776\,{\rm GeV}$ as a reference scale:
\be
t_{\rm tau}=\frac{\tp\ep}{m_{\rm tau}}\approx4\times 10^{-25}\,{\rm s}\,,\qquad \ell_{\rm tau}=\frac{\lp\ep}{m_{\rm mu}} \approx 10^{-16}\,{\rm m}.
\ee
Then,
\ba
\hspace{-1.3cm}&&t_* < 6\times 10^{-28}\,{\rm s}\,,\qquad \ell_*<2\times 10^{-19}\,{\rm m}\,,\qquad E_*>1.0\,{\rm TeV}\,,\qquad \a_0,\a\ll\frac12\,,\\
\hspace{-1.3cm}&&t_* < 10^{-30}\,{\rm s}\,,\qquad \ell_*<3\times 10^{-22}\,{\rm m}\,,\qquad E_*>6 \times 10^{5}\,{\rm GeV}\,,\qquad \a_0=\frac12=\a\,,
\ea
slightly weaker than the bounds from kaons.

%Kaon decay: https://arxiv.org/pdf/hep-ph/9906403.pdf
%http://article.sciencepublishinggroup.com/pdf/10.11648.j.ajpa.s.2015030601.11.pdf

%%%%%%%%%%%%%%%%%%%%%%%%%%%%%%%%%%%%%%%%%%%%%%%%%%%%%%%%%%%%%%%%%%%%%%%%%%%%%%%%%%%%%%%%%

\subsection{\texorpdfstring{$q$}{}-derivatives}
\noindent 
Expression \Eq{qbou} with $\tau=\tau_{\rm tau}$ and $\de\tau=\de\tau_{\rm tau}$ gives the weakest constraint at $\bar\a_0\approx 0.12$ and a tighter one at $\a_0=1/2$:
\ba
\hspace{-1.3cm}&&t_* < 4\times 10^{-19}\,{\rm s}\,,\qquad \ell_*<10^{-10}\,{\rm m}\,,\qquad E_*>2 \times 10^{-6}\,{\rm GeV}\,,\\
\hspace{-1.3cm}&&t_* < 9\times 10^{-21}\,{\rm s}\,,\qquad \ell_*<3\times 10^{-12}\,{\rm m}\,,\qquad E_*>8 \times 10^{-5}\,{\rm GeV}\,,\qquad \a_0=\frac12\,.
\ea
These constraints are stronger than the bounds from the kaon transitions and from the muon lifetime.

%%%%%%%%%%%%%%%%%%%%%%%%%%%%%%%%%%%%%%%%%%%%%%%%%%%%%%%%%%%%%%%%%%%%%%%%%%%%%%%%%%%%%%%%%
%%%%%%%%%%%%%%%%%%%%%%%%%%%%%%%%%%%%%%%%%%%%%%%%%%%%%%%%%%%%%%%%%%%%%%%%%%%%%%%%%%%%%%%%%

\section{Conclusions}\label{concl}
\begin{figure}
\begin{center}
\includegraphics[width=10cm]{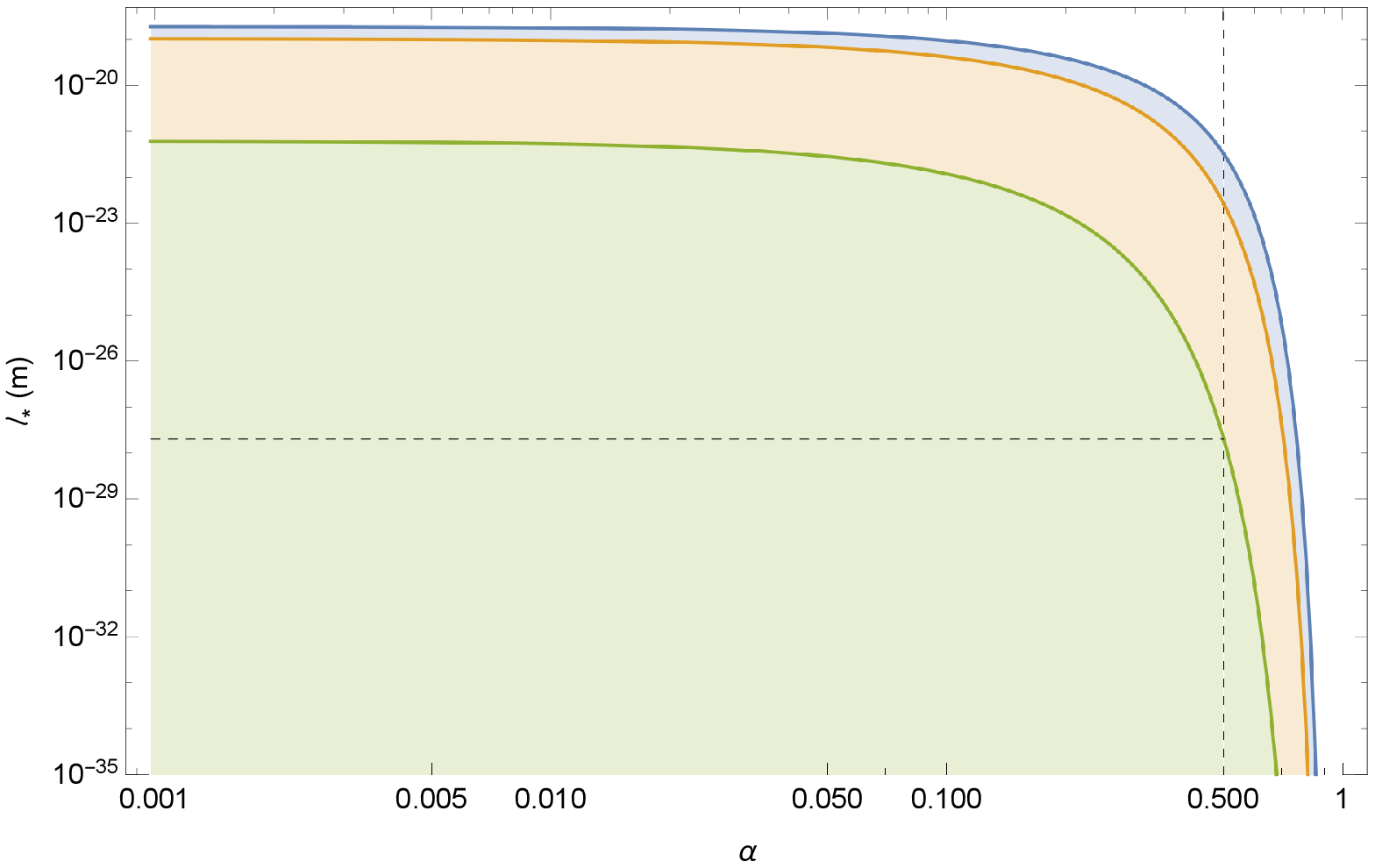}\\
\includegraphics[width=10cm]{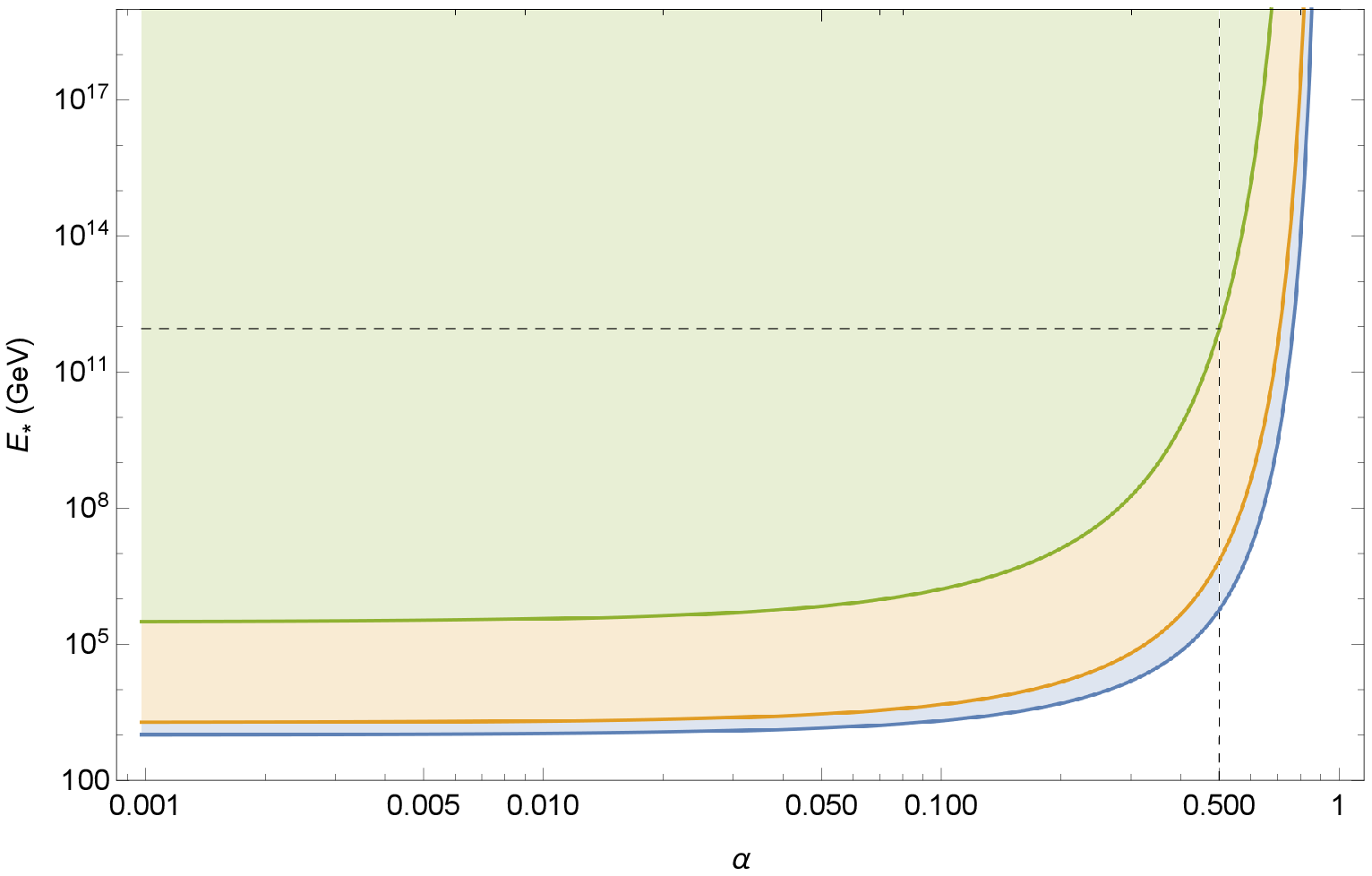}
\end{center}
\caption{\label{fig2} Allowed values of the characteristic scales $\ell_*$ (top) and $E_*$ (bottom) and the fractional exponent $\a$ for the theory with weighted derivatives. Shaded areas correspond to the tau lifetime (blue, outer region), the kaon--antikaon transitions (red, intermediate region) and the muon lifetime bound (green, inner region). The vertical axis is truncated at the Planck scale in both cases.}
\end{figure}

\begin{figure}
\begin{center}
\includegraphics[width=10cm]{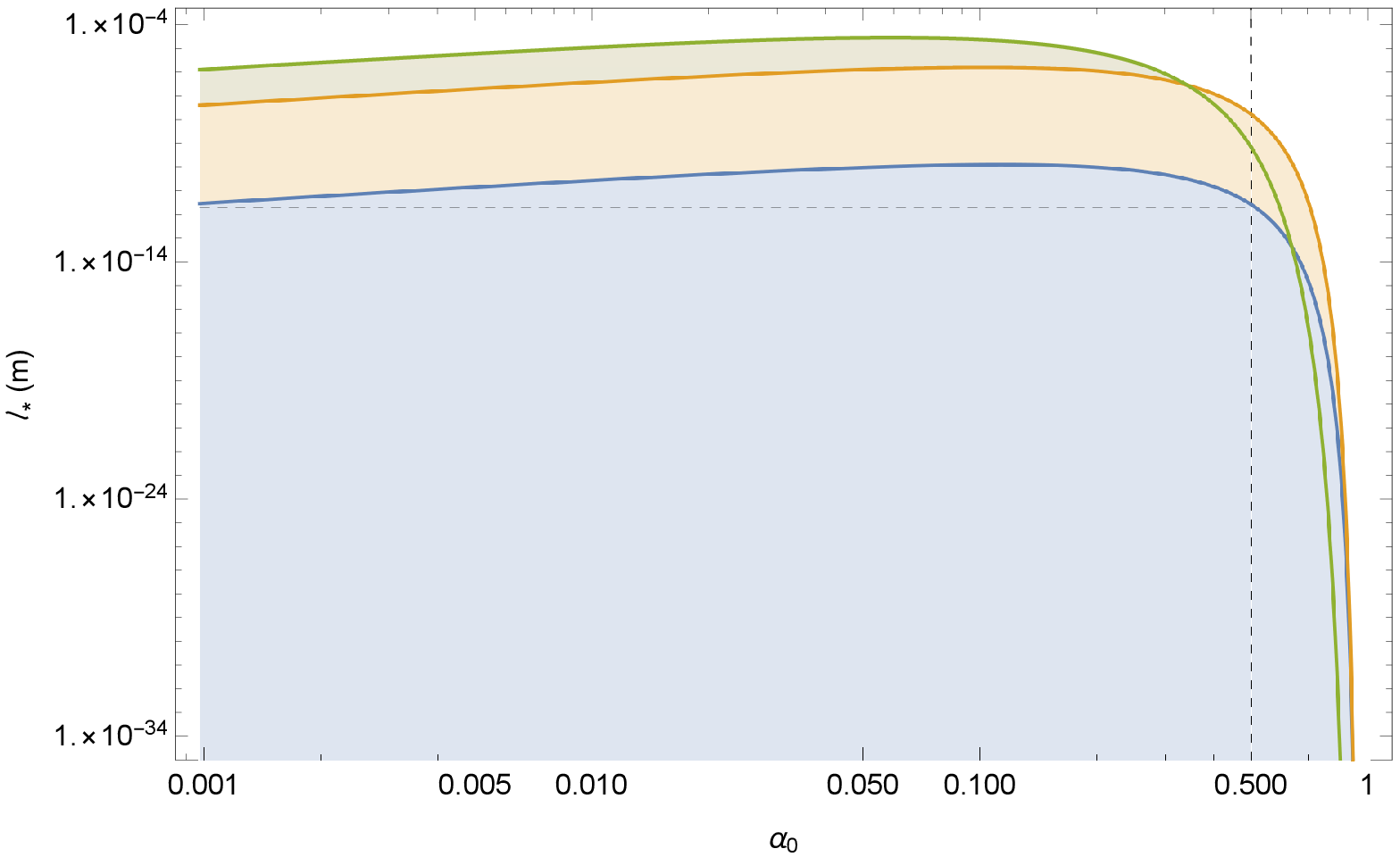}\\
\includegraphics[width=10cm]{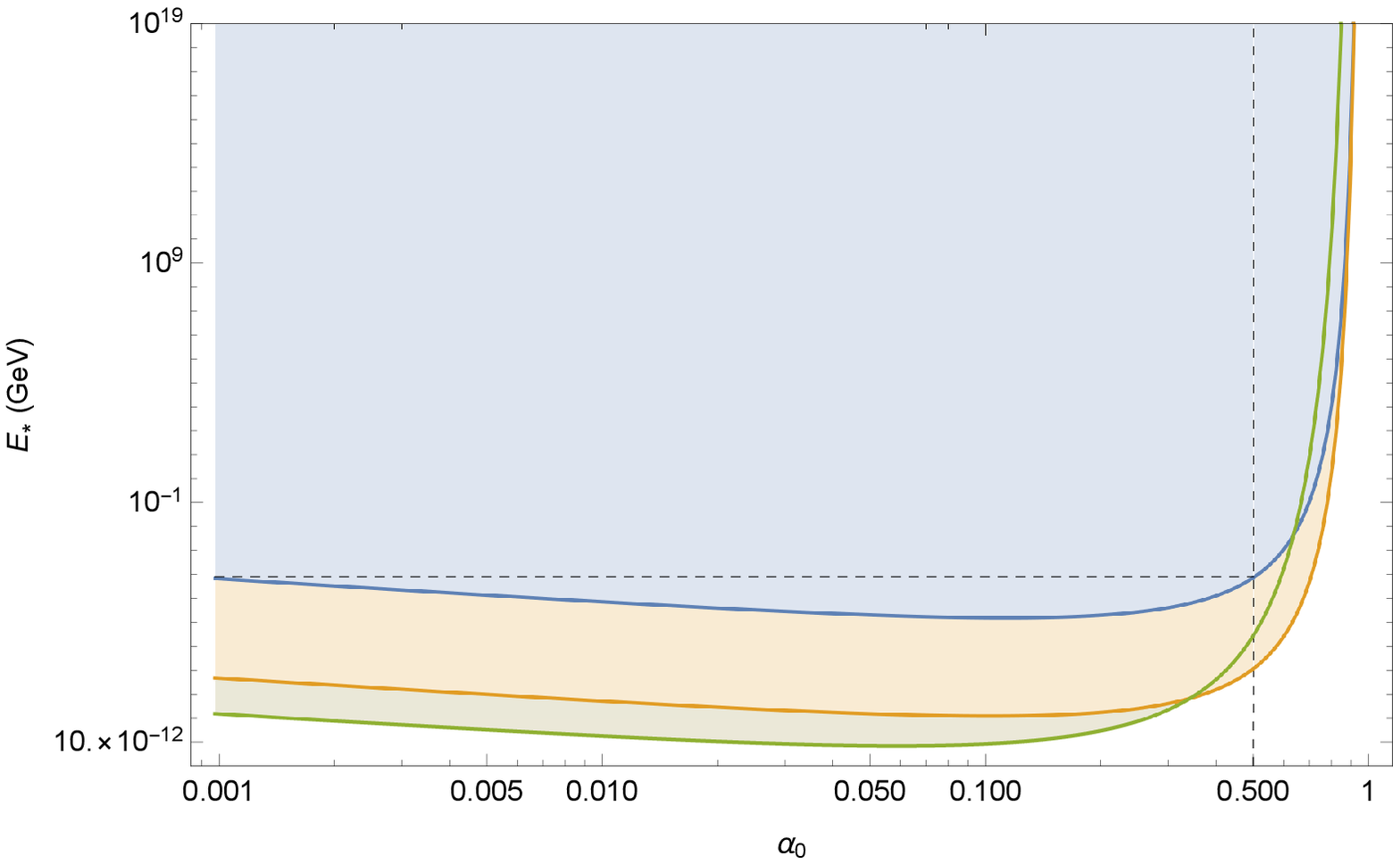}
\end{center}
\caption{\label{fig3} Allowed values of the characteristic scales $\ell_*$ (top) and $E_*$ (bottom) and the fractional exponent $\a_0$ for the theory with $q$-derivatives. Shaded areas correspond to the tau lifetime (blue, inner region), the kaon--antikaon transitions (red, intermediate region) and the muon lifetime bound (green, outer region). The vertical axis is truncated at the Planck scale in both cases.}
\end{figure}

\noindent 
The allowed regions in the $(\a,\ell_*)$ and $(\a,E_*)$ planes from the muon lifetime, the tau lifetime and the kaon mass splitting are showed in Figs.\ \ref{fig2} and \ref{fig3} for the theory with, respectively, weighted and $q$-derivatives, while the bounds at $\a\ll 1$ and $\a=1/2$ are summarized in Table \ref{tab1}.

\begin{table}[t]
\begin{center}
\begin{tabular}{lccc}\hline\hline
Weighted derivatives (absolute)  & $t_*$ (s)           & $\ell_*$ (m) & $E_*$ (GeV) \\\hline
Tau lifetime                     & $<6\times 10^{-28}$ & $<2\times 10^{-19}$ & $>1\times 10^{3}$\\
$K^{0}-\bar{K}^{0}$ transitions  & $<3\times 10^{-28}$ & $<1\times 10^{-19}$ & $>2\times 10^{3}$\\
Muon lifetime                    & $<2\times 10^{-30}$ & $<6\times 10^{-22}$ & $>3\times 10^{5}$\\\hline
Weighted derivatives ($\a=1/2$)  & $t_*$ (s)           & $\ell_*$ (m) & $E_*$ (GeV) \\\hline
Tau lifetime                     & $<1\times 10^{-30}$ & $<3\times 10^{-22}$ & $>6\times 10^{5}$\\
$K^{0}-\bar{K}^{0}$ transitions  & $<9\times 10^{-32}$ & $<3\times 10^{-23}$ & $>7\times 10^{6}$\\
Muon lifetime                    & $<7\times 10^{-37}$ & $<2\times 10^{-28}$ & $>9\times 10^{11}$\\\hline\hline
$q$-derivatives (absolute)       & $t_*$ (s)           & $\ell_*$ (m) & $E_*$ (GeV) \\\hline
Tau lifetime                     & $<4\times 10^{-19}$ & $<1\times 10^{-10}$ & $>2 \times 10^{-6}$\\
$K^{0}-\bar{K}^{0}$ transitions  & $<5\times 10^{-15}$ & $<2\times 10^{-6}$  & $>1 \times 10^{-10}$\\
Muon lifetime                    & $<1\times 10^{-13}$ & $<3\times 10^{-5}$ & $>7 \times 10^{-12}$\\\hline
$q$-derivatives ($\a_0=1/2$)     & $t_*$ (s)           & $\ell_*$ (m) & $E_*$ (GeV) \\\hline
Tau lifetime                     & $<9\times 10^{-21}$ & $<3\times 10^{-12}$ & $>8 \times 10^{-5}$\\
$K^{0}-\bar{K}^{0}$ transitions  & $<5\times 10^{-17}$ & $<2\times 10^{-8}$ & $>1\times 10^{-8}$\\
Muon lifetime                    & $<2\times 10^{-18}$ & $<7\times 10^{-10}$ & $>3 \times 10^{-7}$\\\hline\hline
\end{tabular}
\end{center}
\caption{\label{tab1}Constraints on the multi-fractional theories with weighted and $q$-derivatives from the tau and muon lifetimes and from kaon-antikaon transitions.}
\end{table}

The muon lifetime bounds are the strongest to date for the theory with weighted derivatives. The former strongest bound came from the anomalous magnetic moment of the electron \cite{frc13}, which is $g-2=\tilde\a_\textsc{qed}/\pi$ at one loop in the integer picture. In the fractional picture, $\tilde\a_\textsc{qed}=\a_\textsc{qed}\,v(t)$ \cite{frc8}, and one gets the upper bounds
\[
t_*<\left(\frac{\de\a_\textsc{qed}}{\a_\textsc{qed}}\right)^{\frac{1}{1-\a_0}}t_\textsc{qed}\,,\qquad \ell_*<\left(\frac{\de\a_\textsc{qed}}{\a_\textsc{qed}}\right)^{\frac{1}{1-\a_0}}\ell_\textsc{qed}\,,\qquad E_*>\left(\frac{\a_\textsc{qed}}{\de\a_\textsc{qed}}\right)^{\frac{1}{1-\a_0}}E_\textsc{qed}\,,
\]
where
\[
t_\textsc{qed}=10^{-16}\,{\rm s}\,,\qquad \ell_\textsc{qed}=\frac{\lp t_\textsc{qed}}{\tp}\approx 3\times 10^{-8}\,{\rm m}\,,\qquad E_\textsc{qed}=\frac{\ep\tp}{t_\textsc{qed}}\approx 7\times 10^{-9}\,{\rm GeV}
\]
are the characteristic scale of the quantum electrodynamics processes. Noting that the $2\s$-level relative error is $2\de\a_\textsc{qed}/\a_\textsc{qed}\approx 6.6\times 10^{-10}$, one gets \cite{frc13}
\ba
&&t_* < 7\times 10^{-26}\,{\rm s}\,,\qquad \ell_*<2\times 10^{-17}\,{\rm m}\,,\qquad E_*>10\,{\rm GeV}\,,\qquad \a_0\ll\frac12\,,\label{qed1}\\
&&t_* < 4\times 10^{-35}\,{\rm s}\,,\qquad \ell_*<10^{-26}\,{\rm m}\,,\qquad E_*>2 \times 10^{10}\,{\rm GeV}\,,\qquad \a_0=\frac12\,.\label{qed2}
\ea
Both bounds \Eq{qed1} and \Eq{qed2} are weaker than their analogues \Eq{c1} and \Eq{c2} from the muon lifetime.

Particle-physics bounds on the theory with $q$-derivatives are nowhere nearly as competitive as those on the theory with weighted derivatives, all constraints being about 10 orders of magnitude weaker. In this case, the strongest bound (coming from the tau lifetime) is still considerably weaker than the bound from the Lamb shift \cite{frc12,frc13}, which by itself is not especially compelling ($E_*$ below 1 GeV). In general, in the theory with $q$-derivatives particle physics is less sensitive to multi-scale effects and one must turn to astrophysical observations to reduce the parameter space of the theory significantly \cite{revmu}.

Future directions include the development of the last multi-fractional theory remaining to explore, with fractional derivatives. This proposal, so far not studied in depth because it has no integer frame where calculations simplify, has several features that make it potentially interesting, including an improved renormalizability that the theories with weighted and $q$-derivatives do not have \cite{revmu}. The exploration of all these models, whether renormalizable or not, demonstrates that not all quantum-gravity effects reduce to na\"ive (and unobservable) curvature corrections, and that electroweak and strong processes may have a say in their verification or constraint. Here we saw that dimensional flow, a non-perturbative byproduct of quantum gravitation, can leave an imprint even on quantum perturbative particle phenomena.

%%%%%%%%%%%%%%%%%%%%%%%%%%%%%%%

\section*{Acknowledgments}

\noindent 
A.A.\ and A.M.\ acknowledge support by the NSFC, through the grant No. 11875113, the Shanghai Municipality, through the grant No. KBH1512299, and by Fudan University, through the grant No. JJH1512105. G.C.\ is supported by the I+D grants FIS2014-54800-C2-2-P and FIS2017-86497-C2-2-P of the Ministry of Science, Innovation and Universities.

\medskip

\noindent {\bf Open Access.} This article is distributed under the terms of the Creative Commons Attribution License (\href{https://creativecommons.org/licenses/by/4.0/}{CC-BY 4.0}), which permits any use, distribution and reproduction in any medium, provided the original author(s) and source are credited.

%%%%%%%%%%%%%%%%%%%%%%%%%%%%%%%%%%%%%%%%%%%%%%%%%%%%%%%%%%%%%%%%%%%%%%%%%%%%%%%%%%%%%%%%%
%%%%%%%%%%%%%%%%%%%%%%%%%%%%%%%%%%%%%%%%%%%%%%%%%%%%%%%%%%%%%%%%%%%%%%%%%%%%%%%%%%%%%%%%%

\end{document}